\PassOptionsToPackage{unicode}{hyperref}
\PassOptionsToPackage{hyphens}{url}
\documentclass[
]{article}
\usepackage{xcolor}
\usepackage{amsmath,amssymb}
\usepackage{iftex}
\ifPDFTeX
  \usepackage[T1]{fontenc}
  \usepackage[utf8]{inputenc}
  \usepackage{textcomp} 
\else 
  \usepackage{unicode-math} 
  \defaultfontfeatures{Scale=MatchLowercase}
  \defaultfontfeatures[\rmfamily]{Ligatures=TeX,Scale=1}
\fi
\usepackage{lmodern}
\ifPDFTeX\else
\fi
\IfFileExists{upquote.sty}{\usepackage{upquote}}{}
\IfFileExists{microtype.sty}{
  \usepackage[]{microtype}
  \UseMicrotypeSet[protrusion]{basicmath} 
}{}
\makeatletter
\@ifundefined{KOMAClassName}{
  \IfFileExists{parskip.sty}{%
    \usepackage{parskip}
  }{
    \setlength{\parindent}{0pt}
    \setlength{\parskip}{6pt plus 2pt minus 1pt}}
}{
  \KOMAoptions{parskip=half}}
\makeatother
\usepackage{longtable,booktabs,array}
\usepackage{calc} 
\usepackage{etoolbox}
\AtBeginEnvironment{longtable}{\small}
\makeatletter
\patchcmd\longtable{\par}{\if@noskipsec\mbox{}\fi\par}{}{}
\makeatother
\IfFileExists{footnotehyper.sty}{\usepackage{footnotehyper}}{\usepackage{footnote}}
\makesavenoteenv{longtable}
\ifLuaTeX
  \usepackage{luacolor}
  \usepackage[soul]{lua-ul}
\else
  \usepackage{soul}
\fi
\usepackage{bookmark}
\IfFileExists{xurl.sty}{\usepackage{xurl}}{} 
\hypersetup{
  hidelinks,
  pdfcreator={LaTeX via pandoc}}

\author{}
\date{}

   \usepackage{tipa}
      \usepackage{needspace}
      \usepackage{etoolbox}
\AtBeginEnvironment{longtable}{\small}
\begin{document}

\textbf{
\section{A Situational Speech Synthesizer for Yorùbá:}
}

\textbf{System Design, Phonological Rule Architecture, and Orthographic
Extensions for Contour Tones}

\emph{Kọ́lá Túbọ̀sún, Adédayọ̀ Olúòkun, Hafiz Adéwuyì, and Dadépọ̀ Adérẹ̀mí}

\emph{YorubaName.com}

\vspace{1.5em}
\textsc{ABSTRACT}

\emph{We present TTSYoruba, a rule-based concatenative diphone speech
synthesizer for Yorùbá, deployed online as part of the YorubaName.com
open dictionary of Yorùbá personal names. The system takes tone-marked
Yorùbá text as input and produces audio output by applying a
hand-crafted phonological rule system to a recorded inventory of 651
diphone units spanning five tonal variants of every consonant-vowel
combination in the language. We describe the phonological architecture
of the system in detail, including our tonal file-selection logic, our
treatment of the three-way nasal disambiguation problem (oral /n/,
nasalized vowel, and syllabic nasal), and the derivation of contextual
rising and falling tones from level-tone input. We also present, as an
orthographic contribution, the adoption of the caron (ˇ) and circumflex
(ˆ), which are symbols with prior standing in Yorùbá phonological
transcription, as standard single-vowel contour tone markers, integrated
into the TTS normalization pipeline and the WriteYoruba keyboard input
tool. The system\textquotesingle s performance was evaluated through a
listener study (N=50), with detailed results on Mean Opinion Scores
(MOS) presented in Section 6.}

\emph{\textbf{Keywords:} Yorùbá, text-to-speech, low-resource languages,
diphone synthesis, contour tones, African language NLP, rule-based
synthesis}
\vspace{1.5em}
\section{\texorpdfstring{\textsc{1.
INTRODUCTION}}{1. INTRODUCTION}}\label{introduction}

Yorùbá is one of the few African languages with a standardized writing
system that compulsorily encodes lexical tone directly in its
orthography. High tone uses the acute accent {[}ó{]}, low tone uses the
grave accent {[}ò{]}, and mid tone is usually left unmarked {[}o{]}.
This feature makes Yorùbá unusually tractable for rule-based
text-to-speech (TTS) synthesis: the graphemic input already carries the
phonological information that most TTS systems must infer from
probabilistic models.

In 2015, the Yorùbá Names Project was established to build a free, open
database of Yorùbá personal names with morpheme breakdowns, tone
markings, meanings, and pronunciation guides (Túbọ̀sún, 2020). A core
engineering goal was automated pronunciation. Rather than manually
recording each name\textquotesingle s audio, which is an approach that
does not scale as the database grows through community contribution, the
project required a synthesis engine capable of generating correct
pronunciation for any correctly tone-marked input name.

This paper describes the resulting system, TTSYoruba, and makes the
following contributions:

\begin{itemize}
\item
  A 651-unit diphone corpus covering all phonologically predicted
  syllable-tone combinations in Yorùbá, recorded under consistent
  acoustic conditions.
\item
  A complete tonal file-selection rule system specifying, for every
  grapheme-tone combination, which diphone file is selected and under
  what tonal environment conditions.
\item
  A three-way nasal disambiguation system distinguishing oral /n/,
  nasalized vowels, and syllabic nasals, with the decision logic made
  explicit.
\item
  A new diacritical convention, Unicode caron (ˇ) for rising contour and
  circumflex (ˆ) for falling contour on single vowels, to resolve a
  class of previously unsynthesizable names, with integration into both
  the TTS pipeline and the WriteYoruba keyboard tool.
\end{itemize}

The system is deployed at TTSYoruba.com and integrated into the
YorubaName.com dictionary, where it has generated audio for almost
10,500 name entries since launch.

While components of this system have been in iterative development and
use since 2017, this paper serves as the first formal, comprehensive
documentation of the complete tonal file-selection architecture and the
nasal disambiguation logic, intended to establish these specifications
as a public record and a standard for future Yorùbá speech technology
development.

\section{\texorpdfstring{\textsc{2. BACKGROUND AND RELATED
WORK}}{2. BACKGROUND AND RELATED WORK}}\label{background-and-related-work}

\subsection{2.1 Phonological properties of Yorùbá relevant to
TTS}\label{phonological-properties-of-yoruxf9buxe1-relevant-to-tts}

Yorùbá has three contrastive level tones: high (H), mid (M), and low
(L), with the syllable being the primary tone-bearing unit. Tones are
marked on vowels and, where applicable, on syllabic nasals (e.g. ń, \'{m}).
The language has seven oral vowels (a, e, ẹ, i, o, ọ, u) and eighteen
consonants (b, d, f, g, gb, h, j, k, l, m, n, p, r, s, ṣ, t, w, y).
Syllable structure is predominantly CV, with additional types comprising
standalone vowels (V) and syllabic nasals (N). \emph{(see Awobuluyi,
1994, for a history of orthographic development)}.

The three tones do not behave symmetrically under phonological pressure.
Pulleyblank (2004) demonstrates that High tones are the most stable,
most resistant to deletion and most likely to be retained when tones
compete, while Mid tones are the least stable, most readily lost, and
incapable of participating in lexical contour formation. Low tones are
intermediate. This asymmetry is directly encoded in the file-selection
logic of the TTSYoruba system through a a five-file inventory (\_l, \_m,
\_h, \_f, \_r) that includes rising and falling contour variants for
High and Low tones only. No \_mf or \_mr file exists, and no Mid-tone
syllable triggers contextual rising or falling assignment. That design
reflects the phonological fact that mid tones do not form lexical
contours in Yorùbá.

Beyond the three level tones, phonological processes yield two contour
tone types: a rising tone (L+H), realized when a high-tone syllable
directly follows a low-tone syllable in a tonal sandhi environment; and
a falling tone (H+L), realized when a low-tone syllable directly follows
a high-tone syllable. These contour realizations are predictable from
the surrounding tonal context and are encoded in the diphone inventory
as separate units (files suffixed \_r and \_f respectively). (See also
Ọlátúbọ̀sún 2012, pp. 13--15, for experimental data on the perceptual
salience of rising and falling tones for non-native speakers.)

The use of the caron and circumflex as contour tone symbols has
precedent in Yorùbá phonological scholarship: Olmsted (1951) employs
them in the first major international phonemic analysis of the language,
and Oyètádé (1988) uses them in an autosegmental treatment of Yorùbá
tonal processes. Section 5 describes how this notation, previously
confined to phonological transcription, has been extended to
orthographic practice and integrated into the TTSYoruba synthesis
pipeline.

\subsection{2.2 Prior work on Yorùbá
TTS}\label{prior-work-on-yoruxf9buxe1-tts}

Ọdẹ́jọbí et al. (2007) propose a syllable-based duration model for Yorùbá
TTS prosody. Ìyàndá et al. (2014) describe a rule-based grapheme-to-phoneme and syllabification front-end for Yorùbá TTS, evaluated on 302 words against expert transcription; it does not generate speech. TTSYoruba extends this prior work by providing
a complete, deployed system whose rule architecture is documented in
sufficient detail to support replication. Its syllable inventory distinguishes only high, mid and low tones; contour tones are not addressed.

\subsection{2.3 The case for rule-based synthesis in a low-resource
setting}\label{the-case-for-rule-based-synthesis-in-a-low-resource-setting}

Neural TTS systems have achieved high naturalness on well-resourced
languages but require hundreds of hours of annotated speech data that do
not yet exist for Yorùbá. Rule-based concatenative synthesis requires
only a small diphone corpus, produces interpretable and auditable
behavior, and offers consistent output for the target domain of personal
names, which are typically short, clearly articulable, and lack the
prosodic complexity of connected speech. These properties make
rule-based synthesis a principled choice for the current application. Earlier Yorùbá work made the same choice for the same reason: Ìyàndá et al. (2014) adopted a rule-based G2P front-end explicitly because Yorùbá is resource-scarce.

\section{\texorpdfstring{\textsc{3. DIPHONE CORPUS AND TONAL FILE
ARCHITECTURE}}{3. DIPHONE CORPUS AND TONAL FILE ARCHITECTURE}}\label{diphone-corpus-and-tonal-file-architecture}

\subsection{3.1 Corpus construction}\label{corpus-construction}

All diphones were recorded by the first author in a home studio
environment using an Audio-Technica AT2020+ microphone and Audacity
2.1.1 at 44.1 kHz stereo. Individual diphone segments were cut and
normalized to consistent amplitude levels. The inventory is organized as
audio files named by a phoneme-tone key convention:

{\def\LTcaptype{none} 
\begin{longtable}[]{@{}
  >{\raggedright\arraybackslash}p{(\linewidth - 8\tabcolsep) * \real{0.1496}}
  >{\raggedright\arraybackslash}p{(\linewidth - 8\tabcolsep) * \real{0.1068}}
  >{\raggedright\arraybackslash}p{(\linewidth - 8\tabcolsep) * \real{0.2350}}
  >{\raggedright\arraybackslash}p{(\linewidth - 8\tabcolsep) * \real{0.2137}}
  >{\raggedright\arraybackslash}p{(\linewidth - 8\tabcolsep) * \real{0.2949}}@{}}
\toprule\noalign{}
\begin{minipage}[b]{\linewidth}\raggedright
\textbf{File suffix}
\end{minipage} & \begin{minipage}[b]{\linewidth}\raggedright
\textbf{Tone}
\end{minipage} & \begin{minipage}[b]{\linewidth}\raggedright
\textbf{Environment}
\end{minipage} & \begin{minipage}[b]{\linewidth}\raggedright
\textbf{Example file}
\end{minipage} & \begin{minipage}[b]{\linewidth}\raggedright
\textbf{Example word}
\end{minipage} \\
\begin{minipage}[b]{\linewidth}\raggedright
\_l.wav
\end{minipage} & \begin{minipage}[b]{\linewidth}\raggedright
Low
\end{minipage} & \begin{minipage}[b]{\linewidth}\raggedright
Default; not after H or R tone
\end{minipage} & \begin{minipage}[b]{\linewidth}\raggedright
ba\_l.wav
\end{minipage} & \begin{minipage}[b]{\linewidth}\raggedright
\ul{Bà}bá (bà-)
\end{minipage} \\
\begin{minipage}[b]{\linewidth}\raggedright
\_m.wav
\end{minipage} & \begin{minipage}[b]{\linewidth}\raggedright
Mid
\end{minipage} & \begin{minipage}[b]{\linewidth}\raggedright
Anywhere
\end{minipage} & \begin{minipage}[b]{\linewidth}\raggedright
ba\_m.wav
\end{minipage} & \begin{minipage}[b]{\linewidth}\raggedright
\ul{Ba}lógun (ba-)
\end{minipage} \\
\begin{minipage}[b]{\linewidth}\raggedright
\_h.wav
\end{minipage} & \begin{minipage}[b]{\linewidth}\raggedright
High
\end{minipage} & \begin{minipage}[b]{\linewidth}\raggedright
Default; not after L or F tone
\end{minipage} & \begin{minipage}[b]{\linewidth}\raggedright
de\_h.wav
\end{minipage} & \begin{minipage}[b]{\linewidth}\raggedright
A\ul{dé} (-dé)
\end{minipage} \\
\begin{minipage}[b]{\linewidth}\raggedright
\_f.wav
\end{minipage} & \begin{minipage}[b]{\linewidth}\raggedright
Falling (H→L)
\end{minipage} & \begin{minipage}[b]{\linewidth}\raggedright
After H or R tone; or circumflex mark (â)
\end{minipage} & \begin{minipage}[b]{\linewidth}\raggedright
boo\_f.wav
\end{minipage} & \begin{minipage}[b]{\linewidth}\raggedright
Adé\underline{bọ̀}wálé (-bọ̀-)
\end{minipage} \\
\begin{minipage}[b]{\linewidth}\raggedright
\_r.wav
\end{minipage} & \begin{minipage}[b]{\linewidth}\raggedright
Rising (L→H)
\end{minipage} & \begin{minipage}[b]{\linewidth}\raggedright
After L or F tone; or caron mark (ǎ)
\end{minipage} & \begin{minipage}[b]{\linewidth}\raggedright
wa\_r.wav
\end{minipage} & \begin{minipage}[b]{\linewidth}\raggedright
Adébọ̀\ul{wá}lé (-wá-)
\end{minipage} \\
\midrule\noalign{}
\endhead
\bottomrule\noalign{}
\endlastfoot
\end{longtable}
}

\emph{Table 1. Tonal file-suffix conventions and their environments. The
five-way tonal distinction (\_l, \_m, \_h, \_f, \_r) applies to every CV
diphone in the inventory.}

Open vowels ẹ and ọ are represented by doubling the vowel letter in the
file key: ``bẹ́'' maps to \emph{bee\_h.}wav, ``bọ̀'' maps to
\emph{boo\_l.wav}, and so on. The grapheme `ṣ' and the digraph `sh'
share the same audio files (sha\_, she\_, etc.), as they represent the
same phoneme /\textipa{S}/ in Yorùbá orthography.

    \needspace{12\baselineskip}
\subsection{3.2 Inventory size}\label{inventory-size}
Table 2 summarizes the inventory.
{\def\LTcaptype{none} 
\begin{longtable}[]{@{}
  >{\raggedright\arraybackslash}p{(\linewidth - 6\tabcolsep) * \real{0.58}}
  >{\raggedright\arraybackslash}p{(\linewidth - 6\tabcolsep) * \real{0.12}}
  >{\raggedright\arraybackslash}p{(\linewidth - 6\tabcolsep) * \real{0.16}}
  >{\raggedright\arraybackslash}p{(\linewidth - 6\tabcolsep) * \real{0.14}}@{}}
\toprule\noalign{}
\begin{minipage}[b]{\linewidth}\raggedright
\textbf{Category}
\end{minipage} & \begin{minipage}[b]{\linewidth}\raggedright
\textbf{Base units}
\end{minipage} & \begin{minipage}[b]{\linewidth}\raggedright
\textbf{Tone variants}
\end{minipage} & \begin{minipage}[b]{\linewidth}\raggedright
\textbf{Subtotal}
\end{minipage} \\
\begin{minipage}[b]{\linewidth}\raggedright
CV diphones (18 consonants × 7 vowels)
\end{minipage} & \begin{minipage}[b]{\linewidth}\raggedright
126
\end{minipage} & \begin{minipage}[b]{\linewidth}\raggedright
5 per pair
\end{minipage} & \begin{minipage}[b]{\linewidth}\raggedright
630
\end{minipage} \\
\begin{minipage}[b]{\linewidth}\raggedright
Word-initial standalone vowels
\end{minipage} & \begin{minipage}[b]{\linewidth}\raggedright
7
\end{minipage} & \begin{minipage}[b]{\linewidth}\raggedright
3 (L, M, H)
\end{minipage} & \begin{minipage}[b]{\linewidth}\raggedright
21
\end{minipage} \\
\begin{minipage}[b]{\linewidth}\raggedright
Total recorded units
\end{minipage} & \begin{minipage}[b]{\linewidth}\raggedright
---
\end{minipage} & \begin{minipage}[b]{\linewidth}\raggedright
---
\end{minipage} & \begin{minipage}[b]{\linewidth}\raggedright
651
\end{minipage} \\
\midrule\noalign{}
\endhead
\bottomrule\noalign{}
\endlastfoot
\end{longtable}
}

\emph{Table 2. Summary of the TTSYoruba diphone inventory.}

The five tonal variants per CV pair arise from the three level tones
plus the two contextually conditioned contour realizations. Standalone
vowels take three variants because they occur word-initially and do not
participate in the same post-consonant tonal sandhi environments as CV
syllables.

\section{\texorpdfstring{\textsc{4. PHONOLOGICAL RULE
SYSTEM}}{4. PHONOLOGICAL RULE SYSTEM}}\label{phonological-rule-system}

The rule system converts a tone-marked input string into a sequence of
diphone file keys, which are looked up and concatenated to produce the
audio output. The pipeline applies four modules in sequence: (1) Unicode
normalization, (2) syllabification and nasal disambiguation, (3) tonal
file selection, and (4) diphone concatenation. We describe each in turn.

\subsection{4.1 Unicode normalization}\label{unicode-normalization}

Input text is normalized to Unicode NFC (Canonical Decomposition,
followed by Canonical Composition) before processing, ensuring that
diacritic-bearing characters are represented as precomposed forms
regardless of the input encoding path. This is essential because Yorùbá
text may arrive from different keyboards and applications with different
Unicode normalization states.

\subsection{4.2 Tonal file selection}\label{tonal-file-selection}

The core rule for file selection is straightforward: the file suffix is
determined by the combination of the graphemic tone marking on the
current syllable and the tone of the immediately preceding syllable. The
full rule set is captured in Table 1 above.

There are two additional cases that require elaboration:

\ul{Contextual rising tone:} A grapheme marked high tone (acute accent)
is assigned the rising (\_r) diphone file when it immediately follows a
syllable assigned a low or falling tone. Example: in Adébọ̀wálé, the
syllable \emph{-wá-} follows the falling-tone \emph{-bọ̀}-; the file
selected is \emph{wa\_r.wav}, not \emph{wa\_h.wav.} This captures the
phonological process by which a high tone is phonetically realized as a
rising contour in this environment.\footnote{This rule implements,
  computationally, the output of the MAX{[}H{]} and MAX{[}L{]}
  faithfulness rankings that Pulleyblank (2004) argues govern Yorùbá
  tonal stability: a High tone is preserved as rising rather than lost,
  because losing it entirely would incur a MAX{[}H{]} violation more
  costly than the contour realization.} Abraham (1958: xiii–xiv) describes the same phenomenon as a “mid-high rising-tone” on high-toned verbs following a low tone, noting that it “arises from its position after a low-tone and is not an inherent tone.”

\ul{Contextual falling tone}: A grapheme marked low tone (grave accent)
is assigned the falling (\_f) diphone file when it immediately follows a
syllable assigned a high or rising tone. Example: in Adébọ̀wálé, the
syllable \emph{-bọ̀-} follows the high-tone \emph{-dé-}; the file
selected is \emph{boo\_f.wav.}

Table 3 traces the complete file-selection sequence for the name
Adébọ̀wálé:

{\def\LTcaptype{none} 
\begin{longtable}[]{@{}
  >{\raggedright\arraybackslash}p{(\linewidth - 6\tabcolsep) * \real{0.2358}}
  >{\raggedright\arraybackslash}p{(\linewidth - 6\tabcolsep) * \real{0.1715}}
  >{\raggedright\arraybackslash}p{(\linewidth - 6\tabcolsep) * \real{0.2144}}
  >{\raggedright\arraybackslash}p{(\linewidth - 6\tabcolsep) * \real{0.2744}}@{}}
\toprule\noalign{}
\begin{minipage}[b]{\linewidth}\raggedright
\textbf{Input syllable}
\end{minipage} & \begin{minipage}[b]{\linewidth}\raggedright
\textbf{Preceding tone}
\end{minipage} & \begin{minipage}[b]{\linewidth}\raggedright
\textbf{File selected}
\end{minipage} & \begin{minipage}[b]{\linewidth}\raggedright
\textbf{Tone assignment}
\end{minipage} \\
\begin{minipage}[b]{\linewidth}\raggedright
a-
\end{minipage} & \begin{minipage}[b]{\linewidth}\raggedright
Word-initial
\end{minipage} & \begin{minipage}[b]{\linewidth}\raggedright
a\_m.wav
\end{minipage} & \begin{minipage}[b]{\linewidth}\raggedright
Mid (default initial)
\end{minipage} \\
\begin{minipage}[b]{\linewidth}\raggedright
-dé
\end{minipage} & \begin{minipage}[b]{\linewidth}\raggedright
After mid
\end{minipage} & \begin{minipage}[b]{\linewidth}\raggedright
de\_h.wav
\end{minipage} & \begin{minipage}[b]{\linewidth}\raggedright
High (diacritic)
\end{minipage} \\
\begin{minipage}[b]{\linewidth}\raggedright
-bọ̀-
\end{minipage} & \begin{minipage}[b]{\linewidth}\raggedright
After high
\end{minipage} & \begin{minipage}[b]{\linewidth}\raggedright
boo\_f.wav
\end{minipage} & \begin{minipage}[b]{\linewidth}\raggedright
Falling (low after high → \_f)
\end{minipage} \\
\begin{minipage}[b]{\linewidth}\raggedright
-wá-
\end{minipage} & \begin{minipage}[b]{\linewidth}\raggedright
After falling
\end{minipage} & \begin{minipage}[b]{\linewidth}\raggedright
wa\_r.wav
\end{minipage} & \begin{minipage}[b]{\linewidth}\raggedright
Rising (high after falling → \_r)
\end{minipage} \\
\begin{minipage}[b]{\linewidth}\raggedright
-lé
\end{minipage} & \begin{minipage}[b]{\linewidth}\raggedright
After rising
\end{minipage} & \begin{minipage}[b]{\linewidth}\raggedright
le\_h.wav
\end{minipage} & \begin{minipage}[b]{\linewidth}\raggedright
High (diacritic, confirmed by environment)
\end{minipage} \\
\midrule\noalign{}
\endhead
\bottomrule\noalign{}
\endlastfoot
\end{longtable}
}

\emph{Table 3. File-selection trace for Adébọ̀wálé, illustrating
contextual rising and falling tone assignment.}

\subsection{4.3 Nasal disambiguation}\label{nasal-disambiguation}

The nasal consonant /n/ (and by extension /m/) is the most structurally
ambiguous element in Yorùbá orthography for TTS purposes. It may
function in three distinct phonological roles, each mapping to a
different diphone file:

\begin{enumerate}
\def\labelenumi{\arabic{enumi}.}
\item
  Oral onset: /n/ begins a CV syllable. Example: Adénọ́lá → a-dé-nọ́-lá.
  File: \emph{noo\_h.wav} for ``nọ́''.
\item
  Nasalization marker: /n/ following a vowel nasalizes that vowel and is
  not itself a separate syllable. Example: Sànyà → /sã/-/ja/. File: the
  preceding vowel-consonant pair is replaced by the corresponding nasal
  diphone (e.g. \emph{san\_l.wav}).
\item
  Syllabic nasal: /n/ or /m/ constitutes a complete syllable bearing its
  own tone. Example: Àńjọlá → à-ń-jọ-lá. File: \emph{n\_r.wav} (rising,
  because it follows a low-tone syllable).
\end{enumerate}

The disambiguation rules, derived from systematic analysis of the Yorùbá
name corpus, are as follows:

\textbf{\ul{Rule N1:}} (syllabic, explicit). If /n/ or /m/ carries a
tone diacritic (e.g. ń, ǹ, ḿ, m̀), treat as syllabic. File:
\emph{n\_h.wav}, \emph{n\_l.wav}, \emph{m\_h.wav}, etc.

\textbf{\ul{Rule N2}} (nasal vowel, vowel class). If /n/ is preceded by
the vowels /i/ or /u/ in mid-word position, and followed by a consonant,
treat /n/ as a nasalization marker and assign the nasal diphone to the
preceding vowel-onset pair. Example: Àdùntọ́lá → à-dùn-tọ́-lá, because /u/
precedes /n/ which is followed by /t/. The morphological structure
places the boundary after dùn.

\textbf{\ul{Rule N3}} (oral onset, default). If /n/ is preceded by /o/,
/e/, /ọ/, or /ẹ/ in mid-word position and followed by a vowel, treat /n/
as an oral onset consonant for the following syllable. Example: Adénọ́lá
→ a-dé-nọ́-lá.

\textbf{\ul{Rule N4}} (word-final nasalization). Any word-final /n/
nasalizes the preceding vowel.

The three-way classification underlying these rules (i.e: \emph{n} as
consonant onset, n as syllabic nasal, and \emph{n} as nasality marker on
a preceding vowel) is given a formal linguistic treatment in Ajiboye
(2020), who notes that the convention of using \emph{n} to mark nasal
vowels rather than the diacritic {[}˜{]} was a deliberate decision of
the 1974 Orthography Committee made on practical typing grounds. Abraham (1958: vi, xxii) earlier had resorted to hyphens (\textit{hà-\`{n}-tà}) and colons (\textit{k\`{i}n:n\`{i}un}) to prevent exactly this misreading in print. Rules N1--N4 now addresses these ambiguities, seen here in Table 4 across representative names:

{\def\LTcaptype{none} 
\small
\begin{longtable}[]{@{}
  >{\raggedright\arraybackslash}p{(\linewidth - 6\tabcolsep) * \real{0.1600}}
  >{\raggedright\arraybackslash}p{(\linewidth - 6\tabcolsep) * \real{0.2400}}
  >{\raggedright\arraybackslash}p{(\linewidth - 6\tabcolsep) * \real{0.2000}}
  >{\raggedright\arraybackslash}p{(\linewidth - 6\tabcolsep) * \real{0.4000}}@{}}
\toprule\noalign{}
\begin{minipage}[b]{\linewidth}\raggedright
\textbf{Name}
\end{minipage} & \begin{minipage}[b]{\linewidth}\raggedright
\textbf{Syllabification}
\end{minipage} & \begin{minipage}[b]{\linewidth}\raggedright
\textbf{Rule class}
\end{minipage} & \begin{minipage}[b]{\linewidth}\raggedright
\textbf{Notes}
\end{minipage} \\
\begin{minipage}[t]{\linewidth}\raggedright
Sànyà
\end{minipage} & \begin{minipage}[t]{\linewidth}\raggedright
sã -- ja
\end{minipage} & \begin{minipage}[t]{\linewidth}\raggedright
Nasal vowel
\end{minipage} & \begin{minipage}[t]{\linewidth}\raggedright
/n/ after /a/ before consonant → nasal vowel
\end{minipage} \\ \addlinespace
\begin{minipage}[t]{\linewidth}\raggedright
Sáḿbà
\end{minipage} & \begin{minipage}[t]{\linewidth}\raggedright
sá -- \textsyllabic{\'{m}} -- bà
\end{minipage} & \begin{minipage}[t]{\linewidth}\raggedright
Syllabic nasal
\end{minipage} & \begin{minipage}[t]{\linewidth}\raggedright
/m/ before bilabial → syllabic
\end{minipage} \\ \addlinespace
\begin{minipage}[t]{\linewidth}\raggedright
Sángo\footnote{The place name, which is different from the name of the deity, should ordinarily be written as \emph{Sánńgo}, if accounting for all the variations of \emph{n} in the word were the object of the spelling, but this is rarely the case, due to these occasional arbitrariness in Yorùbá spelling. It's the same with the name ``Ògúndé'' (which is pronounced \emph{Ògún\`{n}dé}).}
\end{minipage} & \begin{minipage}[t]{\linewidth}\raggedright
sã -- \textsyllabic{\textipa{N}} -- go
\end{minipage} & \begin{minipage}[t]{\linewidth}\raggedright
Mixed
\end{minipage} & \begin{minipage}[t]{\linewidth}\raggedright
Nasal vowel + syllabic /\textsyllabic{\textipa{N}}/
\end{minipage} \\ \addlinespace
\begin{minipage}[t]{\linewidth}\raggedright
Àńjọlá
\end{minipage} & \begin{minipage}[t]{\linewidth}\raggedright
à -- \textsyllabic{\'{n}} -- jọ -- lá
\end{minipage} & \begin{minipage}[t]{\linewidth}\raggedright
Syllabic nasal
\end{minipage} & \begin{minipage}[t]{\linewidth}\raggedright
Tone mark on /n/ → \emph{n\_r.wav}
\end{minipage} \\ \addlinespace
\begin{minipage}[t]{\linewidth}\raggedright
Adénọ́lá
\end{minipage} & \begin{minipage}[t]{\linewidth}\raggedright
a -- dé -- nọ́ -- lá
\end{minipage} & \begin{minipage}[t]{\linewidth}\raggedright
Oral /n/
\end{minipage} & \begin{minipage}[t]{\linewidth}\raggedright
/n/ between vowels → oral
\end{minipage} \\ \addlinespace
\begin{minipage}[t]{\linewidth}\raggedright
Ìtànìfẹ́
\end{minipage} & \begin{minipage}[tb]{\linewidth}\raggedright
ì -- tã -- ì -- fẹ́
\end{minipage} & \begin{minipage}[t]{\linewidth}\raggedright
Nasal vowel
\end{minipage} & \begin{minipage}[t]{\linewidth}\raggedright
/an/ at morpheme boundary: ìtàn + ìfẹ́
\end{minipage} \\ \addlinespace
\begin{minipage}[t]{\linewidth}\raggedright
Àdùnọlá
\end{minipage} & \begin{minipage}[t]{\linewidth}\raggedright
à -- dùn -- ọ -- lá
\end{minipage} & \begin{minipage}[t]{\linewidth}\raggedright
Nasal diphone
\end{minipage} & \begin{minipage}[t]{\linewidth}\raggedright
/u/ precedes /n/ → \emph{dun\_f.wav} group
\end{minipage} \\ \addlinespace
\midrule\noalign{}
\endhead
\bottomrule\noalign{}
\endlastfoot
\end{longtable}
}

\emph{Table 4. Nasal disambiguation examples showing the three rule
classes and their outputs.}

An acknowledged limitation is the morpheme-boundary case, like in
\emph{Ìtànìfẹ́} (morphological decomposition: ìtàn + ìfẹ́). Rule N3 as
stated incorrectly parses the second syllable as /ta/ + /ni/ rather than
/tã/. Even a name like \emph{Adùnọlá} (which admits a rare alternative
pronunciation \emph{A-dù-nọ́-lá} in some peculiar instances) presents an
apparent exception. However, the most widely attested form is the nasal
variant, and Rule N3 correctly handles it. Resolving the likely many
more exceptions in this class requires either a morphological parser or
a manually curated exception lexicon. The current system handles such
names by lexical override when they are identified as problematic.

\section{\texorpdfstring{\textsc{5. ORTHOGRAPHIC EXTENSION: ENCODING
CONTOUR TONES ON SINGLE
VOWELS}}{5. ORTHOGRAPHIC EXTENSION: ENCODING CONTOUR TONES ON SINGLE VOWELS}}\label{orthographic-extension-encoding-contour-tones-on-single-vowels}

\subsection{5.1 The problem}\label{the-problem}

The contextual rising and falling tone rules in Section 4.2 handle
contour tones that arise from the tonal environment of a syllable within
a correctly tone-marked word. A separate and more difficult class of
cases arises for personal names whose conventional spellings omit the
vowel gemination that standard Yorùbá orthography would use to represent
a contour tone within a single syllable. The caron and circumflex are
adopted here as normalization targets within the TTS pipeline.

The standard solution is geminated vowel (e.g. vowel doubling), where a
name like ``Nike'' is written with a doubled vowel e.g. ``Níkẹ̀ẹ́'', each vowel carrying one component tone of the rising
contour. Native speakers do not typically use this spelling because
established names circulate in forms that omit this doubling, either
because they entered print before systematic tone-marking conventions
were established, or because their bearers prefer the inherited
spelling. A TTS system that processes names as written cannot synthesize
contour tones for this class without either distorting the spelling or
leaving the tone unrepresented. A fuller linguistic and orthographic
treatment of this problem, including its historical context and the
formal equivalences between contour-marked and geminated forms, is
presented in Túbọ̀sún (2026).

Native speakers who encounter geminated contour orthography outside of
linguistic or pedagogical contexts often perceive it as erroneous or
unusual: in the MOS listening study reported in Section 6, one
participant spontaneously noted that the spelling \emph{Akẹ́kọ̀ọ́}, the
standard geminated form of the word for ``student'', was "off," while
the caron equivalent \emph{Akẹ́kọ̌} elicited no such comment. This
asymmetry illustrates the practical cost of geminated notation: even
when phonologically accurate, it may trigger unfamiliarity responses
that interfere with the perception of correctness.

This problem was identified but unresolved in the first version of this
system. We now present a solution.

\subsection{5.2 Proposed convention}\label{proposed-convention}

We propose encoding contour tones on single vowels using two Unicode
diacritics:

\begin{itemize}
\item
  Caron (ˇ), yielding e.g. ǎ, ě, ẹ̌, ǐ, ǒ, ọ̌, ǔ: to encode a rising (L+H)
  contour on a single vowel.
\item
  Circumflex (ˆ), yielding e.g. â, ê, ệ, î, ô, ộ, û: to encode a falling
  (H+L) contour on a single vowel.
\end{itemize}

These marks are already defined in the Unicode Latin Extended range, are
supported in common fonts, and can be combined with the subscript dot
used for the open vowels ẹ and ọ. They map to the existing \_r and \_f
diphone files respectively, so no new audio recordings are required. The
formal equivalences are:

\begin{quote}
ǎ $\equiv$ àá (rising contour on /a/)

â $\equiv$ áà (falling contour on /a/)

ě $\equiv$ èé (rising contour on /e/)

ê $\equiv$ éè (falling contour on /e/)

ẹ̌ $\equiv$ ẹ̀ẹ́ (rising contour on /ẹ/)

ệ $\equiv$ ẹ́ẹ̀ (falling contour on /ẹ/)

ǐ $\equiv$ ìí (rising contour on /i/)

î $\equiv$ íì (falling contour on /i/)

ǒ $\equiv$ òó (rising contour on /o/)

ô $\equiv$ óò (falling contour on /o/)

ọ̌ $\equiv$ ọ̀ọ́ (rising contour on /ọ/)

ộ $\equiv$ ọ́ọ̀ (falling contour on /ọ/)

ǔ $\equiv$ ùú (rising contour on /u/)

û $\equiv$ úù (falling contour on /u/)
\end{quote}

These equivalences are already reflected in the rulebook that governs
the TTS engine. Entry 1120b, for instance, specifies that the caron form
``rǎn'' (as in Arómọlárǎn) maps to \emph{roon\_r.wav}, exactly as the
geminated form \emph{ràán} does. Entry 261b specifies that ``kẹ̌'' (as in
Adéníkẹ̌) maps to \emph{kee\_r.wav}. The convention therefore requires no
changes to the diphone inventory; only a normalization step in the
preprocessing pipeline.

\needspace{18\baselineskip}
\subsection{5.3 Examples}\label{examples}

{\def\LTcaptype{none} 
\begin{longtable}[]{@{}
  >{\raggedright\arraybackslash}p{(\linewidth - 6\tabcolsep) * \real{0.2036}}
  >{\raggedright\arraybackslash}p{(\linewidth - 6\tabcolsep) * \real{0.2358}}
  >{\raggedright\arraybackslash}p{(\linewidth - 6\tabcolsep) * \real{0.2036}}
  >{\raggedright\arraybackslash}p{(\linewidth - 6\tabcolsep) * \real{0.2529}}@{}}
\toprule\noalign{}
\begin{minipage}[b]{\linewidth}\raggedright
\textbf{Common spelling}
\end{minipage} & \begin{minipage}[b]{\linewidth}\raggedright
\textbf{Full tonal form}
\end{minipage} & \begin{minipage}[b]{\linewidth}\raggedright
\textbf{Contour-marked form}
\end{minipage} & \begin{minipage}[b]{\linewidth}\raggedright
\textbf{TTS file key (contour syllables)}
\end{minipage} \\
\begin{minipage}[b]{\linewidth}\raggedright
Aromọlaran
\end{minipage} & \begin{minipage}[b]{\linewidth}\raggedright
Arómọláràán
\end{minipage} & \begin{minipage}[b]{\linewidth}\raggedright
Arómọlárǎn
\end{minipage} & \begin{minipage}[b]{\linewidth}\raggedright
\emph{roon\_r.wav} (rǎn)
\end{minipage} \\
\begin{minipage}[b]{\linewidth}\raggedright
Ijerisi
\end{minipage} & \begin{minipage}[b]{\linewidth}\raggedright
Ìjẹ́rìísí
\end{minipage} & \begin{minipage}[b]{\linewidth}\raggedright
Ìjẹ́rǐsí
\end{minipage} & \begin{minipage}[b]{\linewidth}\raggedright
\emph{ri\_r.wav} (rǐ)
\end{minipage} \\
\begin{minipage}[b]{\linewidth}\raggedright
Adenikẹ
\end{minipage} & \begin{minipage}[b]{\linewidth}\raggedright
Adéníkẹ̀ẹ́
\end{minipage} & \begin{minipage}[b]{\linewidth}\raggedright
Adéníkẹ̌
\end{minipage} & \begin{minipage}[b]{\linewidth}\raggedright
\emph{kee\_r.wav} (kẹ̌)
\end{minipage} \\
\begin{minipage}[b]{\linewidth}\raggedright
Ajafẹtọ
\end{minipage} & \begin{minipage}[b]{\linewidth}\raggedright
Ajàfẹ́tọ̀ọ́
\end{minipage} & \begin{minipage}[b]{\linewidth}\raggedright
Ajàfẹ́tọ̌
\end{minipage} & \begin{minipage}[b]{\linewidth}\raggedright
\emph{too\_r.wav} (tọ̌)
\end{minipage} \\
\begin{minipage}[b]{\linewidth}\raggedright
Adebiyi
\end{minipage} & \begin{minipage}[b]{\linewidth}\raggedright
Adébíyìí
\end{minipage} & \begin{minipage}[b]{\linewidth}\raggedright
Adébíyǐ
\end{minipage} & \begin{minipage}[b]{\linewidth}\raggedright
\emph{yi\_r.wav} (yǐ)
\end{minipage} \\
\begin{minipage}[b]{\linewidth}\raggedright
Alawiye
\end{minipage} & \begin{minipage}[b]{\linewidth}\raggedright
Aláwìíyé
\end{minipage} & \begin{minipage}[b]{\linewidth}\raggedright
Aláwǐyé
\end{minipage} & \begin{minipage}[b]{\linewidth}\raggedright
\emph{wi\_r.wav} (wǐ)
\end{minipage} \\
\begin{minipage}[b]{\linewidth}\raggedright
Ayotomi
\end{minipage} & \begin{minipage}[b]{\linewidth}\raggedright
Ayọ̀ọ́tómi
\end{minipage} & \begin{minipage}[b]{\linewidth}\raggedright
Ayọ̌tómi
\end{minipage} & \begin{minipage}[b]{\linewidth}\raggedright
\emph{yoo\_r.wav} (yọ̌)
\end{minipage} \\
\begin{minipage}[b]{\linewidth}\raggedright
Oyedeji
\end{minipage} & \begin{minipage}[b]{\linewidth}\raggedright
Oyèédèjì
\end{minipage} & \begin{minipage}[b]{\linewidth}\raggedright
Oyědèjì
\end{minipage} & \begin{minipage}[b]{\linewidth}\raggedright
\emph{ye\_r.wav} (yě)
\end{minipage} \\
\midrule\noalign{}
\endhead
\bottomrule\noalign{}
\endlastfoot
\end{longtable}
}

\emph{Table 5. Eight Yorùbá personal names showing the common spelling,
full geminated tonal form, proposed contour-marked form, and the TTS
diphone file key for the contour-bearing syllable.}

\subsection{5.4 Implementation}\label{implementation}

For keyboard input, the WriteYoruba tool (Túbọ̀sún, 2026; see also the reference mapping in Túbọ̀sún et al., 2026) provides
the caron and circumflex combinations on all Yorùbá vowels, including
the open variants \emph{ẹ} and \emph{ọ}, all described in the companion
orthographic paper.

TTS normalization. A preprocessing normalization step was added to the
TTSYoruba pipeline: before grapheme-to-phoneme conversion, caron-marked
and circumflex-marked vowels are expanded to their geminated
equivalents, after which processing continues normally. For example:

\begin{quote}
Arómọlárǎn → Arómọláràán … → [\emph{roon\_r.wav}]
\end{quote}

Caron normalization is currently active for all vowels. Circumflex
support for the full open-vowel set is in progress. The practical result
is input-form equivalence: whether a user enters Níkẹ̀ẹ́ or Níkẹ̌,
the synthesizer produces the same rising-contour audio output.

\subsection{5.5 Historical note}\label{historical-note}

The historical precedent for this notation, the formal orthographic argument for its adoption, and the full equivalence table across all seven Yorùbá vowels are presented in Túbọ̀sún (2026), which proposes the convention as a revision of the standard Yorùbá orthography and sets out a formal recommendation for its use. The present section documents only the computational implementation within the TTSYoruba pipeline.

\needspace{6\baselineskip}
\textbf{6. EVALUATION}

We conducted a Mean Opinion Score (MOS) listening study to assess the
naturalness and intelligibility of TTSYoruba output across four
structurally distinct name categories. The study also addresses a
question raised by Fájọbí and Akọ́mọláfẹ́ (2019), whose investigation of
anglicised Yorùbá personal names found that bearers systematically apply
Yorùbá tonal patterns even to structurally anglicised forms, suggesting
that tonal accuracy, not merely segmental intelligibility, is the
relevant perceptual criterion for a Yorùbá name synthesiser. This
motivated our inclusion of two separate rating dimensions.

\textbf{6.1 Study design and participants}

The study was administered as a browser-based listening instrument
hosted at
\href{https://www.writeyoruba.com/ttstest}{\ul{https://www.writeyoruba.com/ttstest}}.
Fifty participants were recruited from the YorubaName.com contributor
community and through the project\textquotesingle s social media
channels; all were at least 18 years of age. Data collection ran over a
six-day period (26 June -- 2 July 2026). Demographic information
collected at the outset included age range, regional Yorùbá variety,
self-rated Yorùbá proficiency on a five-point scale, and
listening-device type. Responses were anonymous; participants could
optionally provide an email address solely to prevent duplicate
submissions. Each participant completed a short unscored practice trial
before the study proper.

The sample was strongly weighted toward native and near-native speakers
(43 of 50, 86\%), with the remainder rating themselves as advanced (3),
intermediate (3), or beginner (1). Participants spanned the 18--24
(n=4), 25--34 (n=24), 35--44 (n=14), 45--54 (n=5), and 55--64 (n=3) age
brackets, and reported a broad range of regional varieties including
Standard Yorùbá, Ọ̀yọ́, Èkìtì, Ìjẹ̀bú, Ìbọ̀lọ́, Ìgbómìnà, Oǹdó, and Ẹ̀kó. The
median completion time was 4.8 minutes.

A majority of participants listened via phone speaker (30, 60\%), with
the remainder using earphones (12, 24\%), external speakers (3, 6\%),
laptop speakers (4, 8\%), or over-ear headphones (1, 2\%). As phone
speakers may attenuate acoustic cues relevant to tonal perception, this
distribution likely places a conservative floor on the naturalness
scores relative to a controlled headphone-listening condition.

\textbf{6.2 Stimuli}

The stimulus pool comprised synthesised audio generated by TTSYoruba for
personal names drawn from four categories corresponding to the
system\textquotesingle s principal phonological domains:

\begin{itemize}
\item
  \textbf{Category A (Standard tonal):} names requiring only the
  standard three-level tone file-selection rules and no contextual
  contour assignment (e.g. Kọ́lá, Yẹ́misí, Títílọlá, Bájọmọ etc).
\item
  \textbf{Category B (Syllabic nasal):} names containing syllabic nasals
  bearing their own tone, engaging the nasal disambiguation rules of
  Section 4.3 (e.g. Adébánjọ, Ifátúnmbí, Gbénga, Pọ́nlé, etc).
\item
  \textbf{Category C (Geminated contour):} names with rising or falling
  contour tones represented through vowel gemination in standard
  orthography (e.g. Arómọláràán, Adéníkẹ̀ẹ́, Aláwìíyé, Oníkàán).
\item
  \textbf{Category D (Caron/circumflex contour):} the same underlying
  names as Category C, written using the single-vowel contour notation
  proposed in Section 5 (e.g. Arómọlárǎn, Adéníkẹ̌, Aláwǐyé, Oníkǎn,
  etc). This category directly tests the equivalence claim of Section
  5.2.
\end{itemize}

Each participant heard ten stimuli: two random names from Category A,
two from Category B, two matched C/D pairs (four items, ensuring each
session contained both the geminated and caron forms of the same
underlying name), and two additional items from any category.
Presentation order was randomised per participant, and paired C/D items
were not presented adjacently. An attention-check item (a synthesis of
Kọ́lá) was embedded at a random non-initial, non-final position in each
session; sessions rating it 1 or 2 on both dimensions simultaneously
would have been excluded.

\textbf{6.3 Rating instrument}

For each stimulus, participants rated two dimensions on a five-point
scale following the ITU-T P.800 convention: naturalness (1 = completely
unnatural; 5 = completely natural) and intelligibility (1 =
incomprehensible; 5 = fully intelligible). Rating controls were disabled
until the participant had played the audio, ensuring no item was rated
unheard; participants could replay each stimulus freely, and the replay
count was logged. An optional free-text comment field was provided per
stimulus. Participants were encouraged but not required to use
headphones or earphones.

\textbf{6.4 Results}

Fifty participants provided 500 scored ratings (excluding
attention-check items). All 50 sessions passed the attention check
(100\% pass rate), and mean replay count was low and uniform across
categories (1.7--2.0), indicating that stimuli were generally rated on a
single or second listen. Table 6 reports mean naturalness and
intelligibility by category.
\needspace{14\baselineskip}
{\def\LTcaptype{none} 
\begin{longtable}[]{@{}
  >{\raggedright\arraybackslash}p{(\linewidth - 6\tabcolsep) * \real{0.28}}
  >{\raggedright\arraybackslash}p{(\linewidth - 6\tabcolsep) * \real{0.16}}
  >{\raggedright\arraybackslash}p{(\linewidth - 6\tabcolsep) * \real{0.18}}
  >{\raggedright\arraybackslash}p{(\linewidth - 6\tabcolsep) * \real{0.38}}@{}}
\toprule\noalign{}
\begin{minipage}[b]{\linewidth}\raggedright
\textbf{Name category}
\end{minipage} & \begin{minipage}[b]{\linewidth}\centering
\textbf{Naturalness}
\end{minipage} & \begin{minipage}[b]{\linewidth}\centering
\textbf{Intelligibility}
\end{minipage} & \begin{minipage}[b]{\linewidth}\raggedright
\textbf{Example stimuli}
\end{minipage} \\
\begin{minipage}[b]{\linewidth}\raggedright
A: Standard tonal
\end{minipage} & \begin{minipage}[b]{\linewidth}\centering
4.15 (0.92)
\end{minipage} & \begin{minipage}[b]{\linewidth}\centering
4.60 (0.68)
\end{minipage} & \begin{minipage}[b]{\linewidth}\raggedright
Kọ́lá, Títílọlá, Ọpẹ́yẹmí\ldots{}
\end{minipage} \\
\begin{minipage}[b]{\linewidth}\raggedright
B: Syllabic nasal
\end{minipage} & \begin{minipage}[b]{\linewidth}\centering
3.90 (0.97)
\end{minipage} & \begin{minipage}[b]{\linewidth}\centering
4.48 (0.72)
\end{minipage} & \begin{minipage}[b]{\linewidth}\raggedright
Bánkẹ́, Adébánjọ, Ifátúnmbí\ldots{}
\end{minipage} \\
\begin{minipage}[b]{\linewidth}\raggedright
C: Geminated
\end{minipage} & \begin{minipage}[b]{\linewidth}\centering
4.10 (1.01)
\end{minipage} & \begin{minipage}[b]{\linewidth}\centering
4.52 (0.80)
\end{minipage} & \begin{minipage}[b]{\linewidth}\raggedright
Arómọláràán, Adéníkẹ̀ẹ́\ldots{}
\end{minipage} \\
\begin{minipage}[b]{\linewidth}\raggedright
D: Caron/circumflex
\end{minipage} & \begin{minipage}[b]{\linewidth}\centering
4.14 (1.02)
\end{minipage} & \begin{minipage}[b]{\linewidth}\centering
4.60 (0.69)
\end{minipage} & \begin{minipage}[b]{\linewidth}\raggedright
Arómọlárǎn, Adéníkẹ̌\ldots{}
\end{minipage} \\
\begin{minipage}[b]{\linewidth}\raggedright
\textbf{Overall}
\end{minipage} & \begin{minipage}[b]{\linewidth}\centering
\textbf{4.08 (0.98)}
\end{minipage} & \begin{minipage}[b]{\linewidth}\centering
\textbf{4.55 (0.72)}
\end{minipage} & \begin{minipage}[b]{\linewidth}\raggedright
all categories
\end{minipage} \\
\midrule\noalign{}
\endhead
\bottomrule\noalign{}
\endlastfoot
\end{longtable}
}

\textit{Table 6. Mean Opinion Scores (1--5) by name category; standard
deviations in parentheses. Overall N = 500 scored ratings from 50
participants.}

Intelligibility was uniformly high across all four categories
(4.48--4.60), indicating that TTSYoruba reliably produces understandable
output regardless of phonological structure. Naturalness scores were
more moderate (3.90--4.15), consistent with the known characteristics of
concatenative diphone synthesis. A one-way ANOVA found no significant
differences among the four categories on either dimension (naturalness:
F(3,496)=1.79, p=.148; intelligibility: F(3,496)=0.84, p=.472).
Syllabic-nasal names (Category B) received the lowest scores on both
dimensions, a pattern that was consistent across the collection period
and is discussed below.

The central comparison of this study concerns Categories C and D: the
geminated and caron/circumflex notations for the same underlying
contour-toned names. Because each session presented both members of two
matched pairs, we compared them using paired-samples t-tests over 100
matched observations. Neither dimension showed a significant difference:
naturalness t(99)=−1.03, p=.307, d=0.10; intelligibility t(99)=−1.65,
p=.101, d=0.17. The two notations are therefore perceptually equivalent
in synthesis, supporting the claim of Section 5.2. We note that the
small numerical differences that did obtain favoured the caron form on
both dimensions, but neither approached significance in the full sample.

Qualitative comments reinforce and extend these findings in three
respects. First, the dominant theme across all categories, which
appeared in a substantial share of free-text responses, was
inter-syllable pause duration: participants repeatedly described the
synthesis as ``too slow,'' as pronounced ``at syllabic level and not
natural flow,'' and as resembling ``teaching or textbook'' speech
directed at non-native learners. This reflects a known limitation of
concatenative diphone synthesis and is consistent with the moderate
naturalness scores observed throughout. Second, multiple participants
independently observed for different name pairs that the geminated and
caron forms sounded ``exactly'' alike; one wrote of the caron form
\emph{Lọ́fẹ̌} that it ``sounds exactly like the previous one with the
double ee,'' providing qualitative corroboration of the statistical
equivalence result. Third, and of direct relevance to Section 5.1, one
participant described the geminated spelling \emph{Akẹ́kọ̀ọ́} as ``off''
while rating its caron equivalent \emph{Akẹ́kọ̌} as ``this is it ---
good,'' a spontaneous native-speaker reaction illustrating the
perceptual cost of geminated notation in everyday contexts.

Because a subset of participants attended to text--audio correspondence
(including orthographic form) in addition to audio quality, we conducted
a sensitivity analysis. Ten sessions were identified from their comments
as attending to spelling or tonal-marking correspondence. Excluding
these sessions shifted every category mean by at most 0.10 points and
did not alter the outcome of the C/D equivalence tests (naturalness
p=.783; intelligibility p=.211, n=80 pairs). These sessions are
therefore retained in the reported analysis.

Taken together, the evaluation supports three conclusions. TTSYoruba
produces highly intelligible Yorùbá across all four name categories; its
naturalness is moderate and limited primarily by the inter-syllable
discontinuities characteristic of concatenative synthesis; and the
caron/circumflex notation proposed in Section 5 is perceptually
indistinguishable from the established geminated notation, incurring no
cost in either naturalness or intelligibility. To our knowledge this is
the first published listener evaluation of a Yorùbá text-to-speech
system, and the scores reported here establish a baseline against which
future neural or hybrid Yorùbá synthesis systems may be compared.

\section{\texorpdfstring{\textsc{7. KNOWN LIMITATIONS AND FUTURE
WORK}}{7. KNOWN LIMITATIONS AND FUTURE WORK}}\label{known-limitations-and-future-work}

\subsection{7.1 Morpheme-boundary nasal
errors}\label{morpheme-boundary-nasal-errors}

The nasal disambiguation rules fail for a subset of names at morpheme
boundaries where a nasalized vowel is immediately followed by a
vowel-initial morpheme. ``Ìtànìfẹ́'' (ìtàn + ìfẹ́) is the prototypical
case. Resolution requires morphological segmentation or an expanded
lexical exception list.

\subsection{7.2 Prosodic modeling for multi-word
utterances}\label{prosodic-modeling-for-multi-word-utterances}

The system was designed for single-name synthesis and avoids the
prosodic challenges of connected speech (downstep, downdrift, high-tone
raising; cf. Laniran and Clements, 2003). Extending the system to
sentences requires a separate prosodic model and pitch modification at
synthesis time, beyond the scope of concatenative diphone synthesis. The
rule, in principle, could be extended to sentence-level input, but the
constraint there is the nature of pre-recorded diphone concatenation
itself. Without on-the-fly pitch contour modification, the system cannot
model the gradual pitch declination (downdrift), register-resetting
(downstep), or boundary tone effects that characterize natural connected
Yorùbá speech.

Beyond the current architecture, the phonological rule system documented
in this paper is designed to function as a language-independent
preprocessing layer, creating a direct path toward a hybrid architecture
in which the rule system serves as a text normalization front-end for a
neural TTS model. The caron and circumflex forms are expanded to their
geminated equivalents, nasal disambiguation is resolved, and contextual
contour assignment is applied before the normalized input is passed to a
neural acoustic model. Such an architecture would preserve the
linguistic precision of the rule-based layer while delegating acoustic
modeling, including pitch contour, coarticulation, and prosodic
phrasing, to a learned model trained on a naturalistic Yorùbá speech
corpus such as ÌròyìnSpeech (Ògúnrẹ̀mí et al., 2024). We regard this
hybrid approach as the most promising near-term direction for improving
naturalness in Yorùbá TTS, and the rule documentation provided here is
intended partly to support that future work.

\subsection{7.3 Input literacy
dependency}\label{input-literacy-dependency}

The system requires correctly tone-marked input. Users who cannot
produce tone marks cannot use the system without assistance. These are a
significant population, given limited formal Yorùbá literacy instruction
and adoption. The WriteYoruba keyboard tool partially addresses this
barrier. Integration with an upstream Automatic Diacritic Restoration
(ADR) system (cf. Ọ̀rìfẹ̀ et al., 2020) would address it more fully.

The architecture described here is specific to languages in which
lexical tone is marked obligatorily in the standard orthography. Igbo
and several other Nigerian tonal languages do not mark tone in their
conventional written forms, meaning that a rule-based concatenative
system of this type cannot be directly applied without first solving the
upstream problem of automatic tone restoration from unmarked text. For
such languages, a statistical or neural approach, along with the
construction of a tone-marked Igbo corpus, would be a necessary
prerequisite, rather than an extension of the diphone rule system
described here.

\subsection{7.4 Circumflex completion}\label{circumflex-completion}

Circumflex support for falling contour tones on open vowels (ệ, ộ) is
partially implemented. Full support across all vowel types is planned
for the next release.

\subsection{7.5 Upstep}\label{upstep}

A tonal phenomenon termed "upstep", in which a high tone is realized at
a significantly elevated pitch with a raising effect on contiguous
tones, as in \emph{ìsekúse}, is documented in Ọlátúbọ̀sún (2012, pp.
14--15) but is not currently modeled in the TTSYoruba system.
Incorporating upstep into the rule architecture is a direction for
future work.

\section{\texorpdfstring{\textsc{8.
CONCLUSION}}{8. CONCLUSION}}\label{conclusion}

We have described TTSYoruba, the first publicly deployed speech
synthesizer for Yorùbá, and documented its phonological rule
architecture in sufficient detail to support evaluation and replication.
The system\textquotesingle s diphone file-selection logic, nasal
disambiguation rules, and contour tone handling are all made explicit
here for the first time. The new diacritical convention for contour
tones, which extends the system\textquotesingle s coverage to a class of
names that could not previously be correctly synthesized, is a
contribution that goes beyond this system: it offers any orthographic,
computational, or pedagogical tool working with Yorùbá text a
Unicode-compatible, keyboard-accessible way to represent rising and
falling contour tones on single vowels without altering established
spellings.

The system demonstrates that low-resource language TTS development and
linguistic description are mutually reinforcing: building the
synthesizer required formalizing phonological rules that had not
previously been computationally specified, and the synthesis failures
that emerged from deployment drove the orthographic innovation described
in Section 5. We offer this record of iterative development as a
practical model for researchers working on speech technology for other
low-resource tonal languages.

A planned subsequent version will replace the current single-speaker
(male) diphone inventory with a recording by a female speaker, using an
identical file-naming convention and rule architecture. This
substitution would require no changes to the phonological rule system,
demonstrating the modularity of the acoustic and linguistic layers of
the design.

\section{\texorpdfstring{\textsc{ACKNOWLEDGMENTS}}{ACKNOWLEDGMENTS}}\label{acknowledgments}

The authors thank the contributors to the YorubaName.com lexicographer
community, whose ongoing additions to the database provided the test
cases that drove the rule development described in this paper, and whose
questions about name pronunciation over many years exposed the contour
tone problem that Section 5 addresses; and the 50 participants whose
ratings in the MOS evaluation test helped validate the perceptual
equivalence of geminated and caron/circumflex contour notation.; and Dr.
Túndé Adégbọlá for input.

\textbf{Copyright and Usage:} The architecture and phonological rules
documented herein are the intellectual property of the author. This work
is licensed under a Creative Commons Attribution-NonCommercial 4.0
International License (CC BY-NC 4.0). Any commercial application or
reproduction of this specific rule-based synthesizer architecture
requires express written permission from the author.

\section{\texorpdfstring{\textsc{REFERENCES}}{REFERENCES}}\label{references}

Abraham, R. C. (1958). \textit{Dictionary of modern Yoruba}. University of London Press.
Ajíbóyè, O. (2020). Functions and distribution of n in Yorùbá
orthography. \emph{Ihafa: A Journal of African Studies}, \emph{11}(1),
58--68.

Akinlabí, A. (2004). The sound system of Yorùbá. In N. Lawal, M. N. O.
Sadiku, \& A. Dopamu (Eds.), \textit{Understanding Yoruba Life and Culture} (pp.
453--468). Africa World Press.

Awóbùlúyì, Q. (1994). The development of Standard Yoruba. In I. Fodor \&
C. Hagège (Eds.), \emph{Language reform: History and future} (Vol. VI,
pp. 25--42). Helmut Buske Verlag.

Bámgbóṣé, A. (1965). \emph{Yoruba orthography: a linguistic
appraisal with suggestions for reform}. Ibadan Univ. Press.

Bámgbóṣé, A. (1966). \textit{A Grammar of Yoruba}. Cambridge University Press.

Bámgbóṣé, A. (1983). Yorùbá orthography. In A. Bámgbóṣé (Ed.),
\textit{Orthographies of Nigerian Languages}. National Language Centre.

Eberhard, D. M., Simons, G. F., \& Fennig,
C. D. (Eds.). (2020) Yoruba. In \textit{Ethnologue: Languages of the World }(23rd ed.). SIL
International.

Fájọbí, E., \& Akọ́mọláfẹ́, B. (2019). Investigating the phonological
processes involved when Yoruba personal names are anglicized.
\emph{English Language and Literature Studies}, \emph{9}(1), 24--37.
\href{https://doi.org/10.5539/ells.v9n1p24}{\ul{https://doi.org/10.5539/ells.v9n1p24}}

Ìyàndá, A. R., Ọdẹ́jọbí, O. A., Sóyọyè, F. A., \& Akinadé, O. O. (2014). Development of grapheme-to-phoneme conversion system for Yorùbá text-to-speech synthesis. \textit{INFOCOMP Journal of Computer Science, 13}(2), 44–53.

Láníran, Y. O., \& Clements, G. N. (2003). Downstep and high raising:
Interacting factors in Yorùbá tone production. Journal of Phonetics,
31(2), 203--250.

Ọdẹ́jọbí, O. A., Wong, S. H. S., \& Beaumont, A. J. (2007). A fuzzy
decision tree-based duration model for Standard Yorùbá text-to-speech
synthesis. \emph{Computer Speech and Language}, \emph{21}(2), 325--349.
\href{https://doi.org/10.1016/j.csl.2006.06.005}{\ul{https://doi.org/10.1016/j.csl.2006.06.005}}

Ọlátúbọ̀sún, K. (2012). \emph{Studies of initial tonal acquisition by
American English speakers learning Yoruba} {[}Master\textquotesingle s
thesis, Southern Illinois University Edwardsville{]}. Zenodo.
\href{https://doi.org/10.5281/zenodo.20833023}{\ul{https://doi.org/10.5281/zenodo.20833023}}

Olmsted, D. L. (1951). The phonemes of Yoruba. \emph{WORD}, \emph{7}(3),
245--249.
\href{https://doi.org/10.1080/00437956.1951.11659409}{\ul{https://doi.org/10.1080/00437956.1951.11659409}}

Orife, I., Adelani, D. I., Fasubaa, T., Williamson, V., Oyewusi, W. F., Wahab, O., \& Tubosun, K. (2020). \textit{Improving Yorùbá diacritic restoration} (arXiv:2003.10564).
\href{https://doi.org/10.48550/arXiv.2003.10564}{\ul{https://doi.org/10.48550/arXiv.2003.10564}}

Pulleyblank, D. (2004). A note on tonal markedness in Yoruba.
\emph{Phonology}, \emph{21}(3), 409--425.
\href{https://doi.org/10.1017/S0952675704000326}{\ul{https://doi.org/10.1017/S0952675704000326}}

Túbọ̀sún, K. (2020). YorubaName.com: An open Yorùbá names database.
\href{https://yorubaname.com}{\ul{https://yorubaname.com}}

Túbọ̀sún, K. (2026). Marking contour tones in Yorùbá: A typographic and computational proposal. \href{(arXiv:2609.38627).} arXiv. {\ul{https://doi.org/10.48550/arXiv.2609.38627}}

Túbọ̀sún, K., Le Guen, L., Morais, L., \& Williamson, V. (2026). \textit{Yorùbá contour-tone vowels: A reference keyboard mapping} (Version 1.0) [Data set]. Zenodo. \href{https://doi.org/10.5281/zenodo.23305200}{https://doi.org/10.5281/zenodo.23305200}

Zen, H., Tokuda, K., \& Black, A. W. (2009). Statistical parametric
speech synthesis. Speech Communication, 51(11), 1039--1064.

\end{document}